# Electronic spin transport and spin precession in single graphene layers at room temperature

Nikolaos Tombros, Csaba Jozsa, Mihaita Popinciuc, Harry T. Jonkman and Bart J. van Wees

Physics of Nanodevices and Molecular Electronics, Zernike Institute for Advanced Materials, Nijenborgh 4, 9747 AG Groningen, The Netherlands

**Electronic transport in single or few layers of graphene is the subject of intense interest. The specific band structure of graphene, with its unique valley structure and Dirac neutrality point separating hole states from electron states has led to the observation of new electronic transport phenomena such as anomalously quantized Hall effects, absence of weak localization and the existence of a minimum conductivity[1]. In addition to dissipative transport also supercurrent transport has already been observed[2]. It has also been suggested that graphene might be a promising material for spintronics and related applications, such as the realization of spin qubits, due to the low intrinsic spin orbit interaction, as well as the low hyperfine interaction of the electron spins with the carbon nuclei[3,4]. As a first step in the direction of graphene spintronics and spin qubits we report the observation of spin transport, as well as Larmor spin precession over micrometer long distances using single graphene layer based field effect transistors. The "non-local" spin valve geometry was used, employing four terminal contact geometries with ferromagnetic cobalt electrodes, which make contact to the graphene sheet through a thin oxide layer. We observe clear bipolar (changing from positive to negative sign) spin signals which reflect the magnetization direction of all 4 electrodes, indicating that spin coherence extends underneath all 4 contacts. No significant changes**

**in the spin signals occur between 4.2K, 77K and room temperature. From Hanle type spin precession measurements we extract a spin relaxation length between 1.5 and 2 μm at room temperature, only weakly dependent on charge density, which is varied from *n*~0 at the Dirac neutrality point to *n* = 3.6 $10^{16}$/m$^2$. The spin polarization of the ferromagnetic contacts is calculated from the measurements to be around 10%.**

The elementary device geometry to detect spin transport is a (two-terminal) spin valve where a non-magnetic medium is contacted by two ferromagnetic electrodes[5]. Hill *et al.*[6] have studied devices with 200 nanometer spaced transparent permalloy contacts to a single graphene layer. Relatively large magnetoresistances of several hundred Ω were observed in the magnetic field region where the magnetization direction switched. However no clear distinction between parallel and anti-parallel configurations could be observed. These experiments suggest the possibility of spin dependent transport. However, the potential contribution of spurious effects has to be excluded by other experiments.

In our experiments we adopt the four terminal "non-local" technique[7,8]. Here the charge current path can be fully separated from the voltage detection circuit. As explained in Fig. 1, if spin injection and spin transport takes place, a bipolar spin signal should be observed which changes sign when the magnetization configuration of the electrodes switches from parallel to anti-parallel. Due to the absence of a background resistance, the non-local technique is less sensitive to device resistance fluctuations and spurious magnetoresistances (such as Hall effects), as compared to the standard two-terminal spin valve technique. The non-local technique has been applied to metals[7,8], semiconductors[9] and carbon nanotubes[10].

For reasons described in the Supplementary Information, we failed to observe spin transport in graphene devices with transparent low Ohmic ferromagnetic contacts. Crucial for our experiments is the use of tunnel barriers between the ferromagnetic electrodes and the graphene layer underneath, in order to increase the spin dependent interface resistance, and combat the conductivity mismatch problem[11,12,13]. Here the presence of a thin $Al_2O_3$ layer should (at least in principle) create a spin dependent tunnel barrier. Also, it is essential for our devices with electrodes which completely overlap the graphene strips that the barrier should be transparent enough to result in measurable (<1M Ω) resistances, but at the same time opaque enough so that carriers can pass underneath it with conservation of spin direction (see Fig. 1c,d). Our measurements show that we have actually achieved this. However, given the expected thickness of the oxide of 0.8 nm it is unlikely that the tunnel barrier will be uniform, and pinholes may be expected.

The fabrication of the devices is described in the Methods section. Fig. 1a shows the device geometry and Fig. 1b the measurement setup. We prepared cobalt electrodes with widths from 80 nm to 800 nm. Their coercive fields depend on their width and range from 150 mT down to 30 mT. Although there is some scatter in the coercive fields for similar contact widths, we can always reconstruct which contact is switching at what magnetic field from the measurements of the various spin valve signals[14]. The non-local resistance is recorded while the magnetic field (applied in the *y*-direction) is swept from a negative value to a positive value followed by an opposite sweep back to negative values. The electrical measurements are performed with standard low frequency lock-in techniques. Currents are in the range 100 nA to 5 µA.

Fig. 2a shows a typical non-local measurement of device #1 at 4.2K. Since there are 4 ferromagnetic electrodes involved, several resistance values are observed which we can associate with a particular configuration of magnetization directions. We note that spin transport over at least 330 nm (the spacing between the centre electrodes) can occur. However, the observation that there are at least 3 resistance levels indicates that also the outer electrodes are relevant as spin injectors and detectors implying an even longer length scale of 1 μm. This device was the only one sofar where we could also observe a spin valve signal in the two-terminal "local" geometry (Fig. 2b). Here the resistance is dominated by the contact resistances which show fluctuations as a function of time, resulting in a noise background. Nevertheless at the switching fields of the central electrodes 2 and 3 an increase of typically 60 Ω can be observed. This is in agreement with the fact that, if both can be measured, the local signal should be larger than the non-local signal[15]. Note also that the identification of these switches as spin signals would have been very difficult without the support of the non-local measurements. Fig. 2c shows a comparison between 4.2K and 77K data of device #1. Here the magnetization direction of only one of the central electrodes is reversed. This results in a typical "minor loop" shape, known from measurements on spin valves with electrodes with different coercivities. We observe only a weak dependence on temperature.

To investigate the dependence of the signal on the electrode separation at 77K we prepared and measured devices #2 and #3 where graphene strips with 2 μm width are used. On each device a sequence of ferromagnetic contacts with monotonically increasing spacings and different widths were deposited. Details will be given elsewehere (Jozsa, C. *et al.*, in preparation). We summarize the spin signals as follows: spacing 350 nm: 31 Ω,  550 nm: 35 Ω, 1 μm: 21 Ω, 1.5 μm: 21 Ω, 2 μm: 22 Ω,  3 μm:  8 Ω, and  5 μm: 5 Ω. This shows that the spin signals start to go down at spacings beyond 2 μm, indicating a spin relaxation length of about 2 μm.

The observation of spin transport at room temperature is shown in Fig. 3a for device #4. The magnitude of the spin signal (~6 Ω) is comparable to that observed at 77K for a 3 μm spacing device. Fig. 3b shows the gate voltage dependence of the non-local signal in parallel and antiparallel configurations. In agreement with the principle of the non-local technique it shows that the parallel configuration always yields a positive value and the antiparallel a negative one. In Fig. 3c the gate voltage dependence of the graphene resistance is plotted. The typical shape for single graphene layers is observed, with a slightly shifted neutrality point at $V_g$ =19 V , where the resistivity reaches a maximum of 3.2 k$\Omega$[1]. A comparison between Fig. 3c and Fig. 3b shows that there is only a small decrease of the spin signal at the Dirac point.

Finally we present room temperature Hanle type spin precession experiments. For this purpose first a magnetic field is applied in the *y*-direction to prepare the electrodes in a parallel or antiparallel magnetization direction. Then this field is removed and a B-field in the *z*-direction is scanned. The data is shown in Fig. 4. Electrons are injected with a spin polarization in the up direction by contact 3. They precess around the *B*-field with a Larmor frequency $\omega_L = g\,\mu_B\,B/\hbar$, with *g* the effective Landé-factor (~2) and $\mu_B$ the Bohr magneton. The data shows that spin-up electrons injected by contact 3 precess while they are diffusing towards contact 2. At *B*~100 mT their average precession angle when they arrive at this contact is 180 degrees, resulting in a sign reversal of the spin signal.

Since the spin density is constant along the *y*-direction, the detailed shape of the data was fitted with the 1-dimensional Bloch equations which describe the combined effect of diffusion, precession and spin relaxation[7,8,9]. From the fitting procedure both the diffusion constant *D* and the transverse spin relaxation time *T2* can be obtained. As we will see later, the longitudinal relaxation time *T1*, extracted from the exponential length dependence of the spin valve signal in sample #5, has similar values with *T2* and therefore the use of a single spin relaxation time $\tau_{sf}$ is justified. The fits indicate a spin relaxation length, $\lambda_{sf} = \sqrt{D \tau_{sf}}$, between 1.5 to 2 μm, with only minor differences between the high density regime and the Dirac point.

We now calculate $D$ from the measured conductivity $\sigma$. At an energy $\varepsilon$ sufficiently far away from the Dirac point the density of states for a single graphene layer is given by $\nu(\varepsilon) = g_v g_s 2\pi |\varepsilon| / (h^2 v_F^2)$, which includes the two-fold valley ($g_v = 2$) and spin ($g_s = 2$) degeneracy ($v_F \sim 10^6$ m/s)[16]. Integration yields the total density $n(\varepsilon) = g_v g_s \pi \varepsilon^2 / (h^2 v_F^2)$. From the measurements of Fig. 3c, together with the device geometry ($L = 3$ μm and $W = 600$ nm) we obtain the conductivity $\sigma = 1.2 \cdot 10^{-3}/\Omega$ at $V_g = -40$ V where $n = 3.6 \ast 10^{16}/\text{m}^2$.

The Einstein relation reads $\sigma = \nu e^2 D$, where the diffusion constant in two-dimensions is given by $D = \frac{1}{2} v_F l$, with $l$: the scattering mean free path. We find $D = 1.8 \cdot 10^{-2}$ m$^2$/s and $l = 36$ nm. This value is very close to the value of $D$ obtained by independent fit of the data in Fig. 4. This supports the interpretation of the signal as being due to the the Larmor precession of diffusing electron spins.

We use a similar procedure at the Dirac neutrality point ($V_g = 19$V). Here we start with an expression given by Refs. 4, 17: $\nu(\varepsilon=0) = 4\pi / (h v_F l)$. Applying the Einstein relation with $\sigma(V_g = 19\text{V}) = 0.33 \cdot 10^{-3}/\Omega$ we find $D = 2.2 \cdot 10^{-2}$ m$^2$/s. Given the numerical uncertainties, this shows that the diffusion constant is not changed significantly compared to the value at high metallic densities. Although the agreement is less good as at higher densities the comparison with the fitted value in Fig. 4 confirms that the description of simultaneous spin diffusion and precession also applies at the Dirac neutrality point.

The obtained values for the spin signals can be now compared with theory. We modify the three-dimensional description of ref. 8 for use in two-dimensions:

$$R_{non-local} = \frac{P^2 \lambda_{sf}}{2W\sigma} \exp(-L/\lambda_{sf}) \qquad (1)$$

Here $W$ is the width of the graphene strip, $L$ the spacing of the central electrodes, and $P$ is the spin polarization of the contacts. Inserting the obtained values of the measured spin signals at room temperature the formula gives $P \sim 0.1$.

Finally, we studied device #5 (graphene width: 1 µm) and we found for the dependence of the spin signals on the spacing $L$: $L$=1.5 µm: 2.0 Ω, 3 µm: 0.8 Ω, 5 µm: 0.25 Ω. These measurements are consistent with $\lambda_{sf}$ in the range 1.5 – 2 µm and a longitudinal relaxation time ($T1$) of the same order as $T2$ obtained previously.

In conclusion we demonstrated all electrical spin injection and detection in the two dimensional carrier system formed by a single graphene layer. Spin transport is found to be relatively insensitive to the temperature. The use of the non-local technique together with artificial tunnel barriers has allowed us to observe spin transport. The contacts most likely consist of some relatively transparent regions which dominate the transport from the ferromagnet into the graphene. Optimization of these contacts is expected to result in even larger spin signals.

We believe that the observed spin relaxation length and relaxation times in our devices are limited by extrinsic impurity scattering and not by the intrinsic properties of graphene. The mobility of our devices is about 2000 cm$^2$/Vs. This is not exceptionally low[1], but improved fabrication techniques should allow for an increase of the scattering time, which will increase both $D$ and $\tau_{sf}$, and thus also $\lambda_{sf}$. This will make room temperature spin transport possible over even longer distances, probably ultimately limited by electron-phonon scattering. At low temperatures the reduction of the role of spin-orbit interaction by the combination of high mobility graphene layers together with quantum confinement should make it possible to increase the spin relaxation times considerably.

**Methods**

The devices (Fig. 1a) are prepared on a doped silicon substrate, insulated by a 500 nm thick silicon oxide. A contact to the substrate allows for a gate voltage to control the charge density in the graphene layer. By means of electron beam lithography a Ti/Au marker pattern is deposited. Next graphene flakes are deposited on the substrate using HOPG (Advanced Ceramics) by a repeated peeling technique[1]. The identification of the single graphene flakes is done by a combination of optical and atomic force microscopy. Sufficiently thin graphene flakes (<0.5 nm) and suitable sizes are selected and their positions relative to the markers are recorded. Next a 0.6 nm thick layer of Al is deposited on the substrate under ultra high vacuum conditions. The substrate is cooled to 77K, to increase the uniformity of the Al layer. The layer is oxidized in a 100 mbar $O_2$ pressure for 1 hour and taken out to ambient atmosphere. Electrical measurements show that there is no conductance through this layer, indicating that the Al has been fully oxidized into $Al_2O_3$. Finally, using electron beam lithography the ferromagnetic contacts (50 nm thick) are fabricated. The devices are then either cooled down to 4.2 K or 77 K (devices #1, #2, #3) or measured at room temperature (devices #4 and #5) all in vacuum.

We measured contact resistances ranging between 5 k$\Omega$ and 50 k$\Omega$, with a few above 100 k$\Omega$. For example, the contact resistance of contact 2 is measured by passing current from contact 1 to 2, and measuring the voltage between 2 and 3. No clear scaling was found between the width of the contacts and their resistances, indicating some form of contact inhomogeneity. The gate voltage dependence of graphene on these devices is similar to that shown in Fig. 3c.

**Supplementary Information** accompanies the paper on www.nature.com/nature

**Acknowledgements** We thank B. Wolfs, S. Bakker, E. Koop, A. Morpurgo, R. Pandian, G. Palasantzas and N. Katsonis for technological support and discussions. We acknowledge support of NWO (PIONIER grant) and MSC[plus] for financial support.

**Author Information** The autors declare no competing financial interests. Correspondence and requests for materials should be addressed to N.Tombros (N.Tombros@rug.nl)

**Figures**

**Figure 1|Spin transport in a four terminal spin valve device. a,** SEM micrograph of a four terminal single layer graphene spin valve. Cobalt electrodes are evaporated across a single layer graphene strip prepared on a SiO$_2$ surface. **b,** The non-local spin valve geometry: A current *I* is injected from electrode 3 through the Al$_2$O$_3$ barrier into graphene and is extracted at contact 4. The voltage difference is measured between contact 2 and 1. The non-local resistance is: $R_{non-local}=(V_+-V_-)/I$. **c,** Illustration of spin injection and spin diffusion for electrodes having parallel magnetizations. Injection of up spins by contact 3 results in an accumulation of spin-up electrons underneath contact 3, with a corresponding deficit of spin-down electrons. Due to spin relaxation the spin density decays on a scale given by the spin relaxation length. The dots show the electric voltage measured by contact 1 and 2 in the ideal case of 100% spin selectivity. A positive non-local resistance is measured. (Note that a larger positive signal can be obtained by reversing the magnetization direction of contact 1). **d,** Spin injection and spin diffusion for antiparallel magnetizations. The voltage contacts probe opposite spin directions resulting in a negative non-local resistance.

**Figure 2| Spin transport at 4.2K and 77K. a,** Non-local spin valve signal for device #1 at 4.2K. The sweep directions of the magnetic field are indicated (red/green arrows). The magnetic configurations of the electrodes are illustrated for both sweep directions. The widths of the electrodes are 1: 330 nm, 2: 90 nm, 3: 140 nm, 4: 250 nm, and the electrode spacings 1-2, 2-3, 3-4: 330 nm. The graphene width is 1.4 µm. **b,** A two terminal local spin valve signal (measured between contacts 2 and 3) of about 60 Ω is measured at 4.2K. **c,** Spin signals at 4.2K and 77K. A "minor loop" is

observed because the magnetic field sweep is reversed before contact 2 has reversed its magnetization direction.

**Figure 3|Spin transport at room temperature. a,** Non-local spin valve signal at room temperature for device #4. The magnetic field sweep directions are indicated as well as the magnetic configurations. The electrode widths are 1: 500 nm  2: 150 nm,  3: 90 nm,  4: 400 nm and electrode spacings 1-2, 3-4:  450 nm,  2-3: 3 µm. The graphene width is 600 nm. **b,** Gate voltage dependence of the spin signals in parallel and anti-parallel directions. The spin signal has a weak minimum at the Dirac point. **c,** Gate voltage dependence of the graphene resistance. The maximum resistivity is 3.2 kΩ at the Dirac neutrality point ($V_g$ = 19V).

**Figure 4|Hanle spin precession.** Hanle spin precession in the non-local geometry (device #4) measured as a function of the perpendicular magnetic field $B_z$ for **a,** parallel and **b,** antiparallel configurations. By application of a magnetic field in the y-direction the device is first prepared in the antiparallel/parallel configuration (see inset). A comparison is made between the data obtained at the Dirac point ($V_g$=19V) and at $V_g$ = -40V. The solid lines are fits with the 1-dimensional Bloch equations. The obtained parameters are shown in the insets, together with the corresponding spin relaxation lenghts. Note that the signal in (**a**) is much smaller compared to (**b**), due to a sudden change in the properties of the 150 nm wide contact between the measurements.

**Figure 1:**

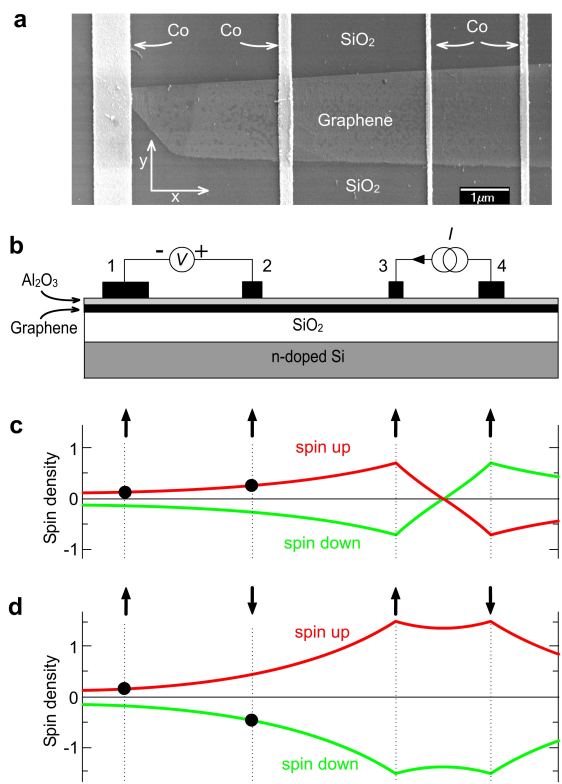

**Figure 2:**

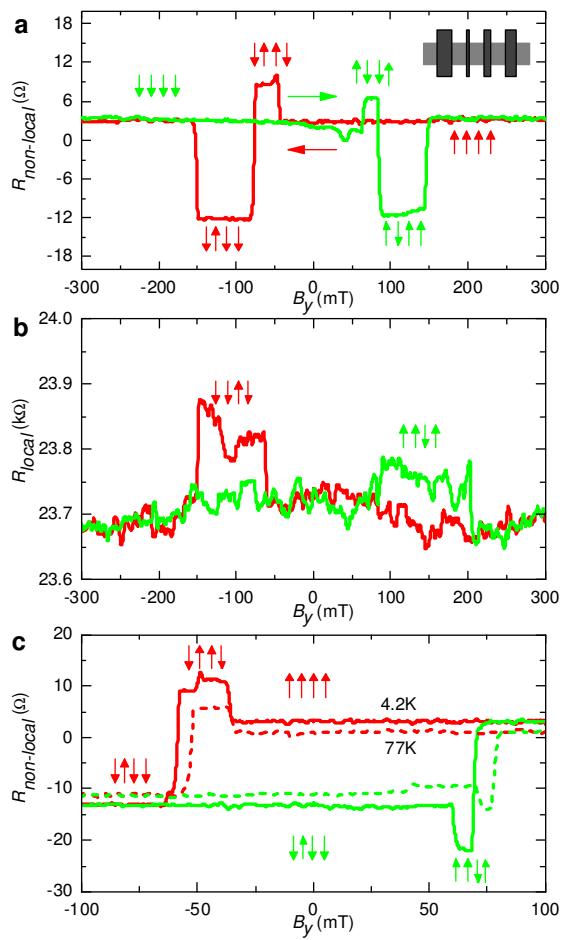

**Figure 3:**

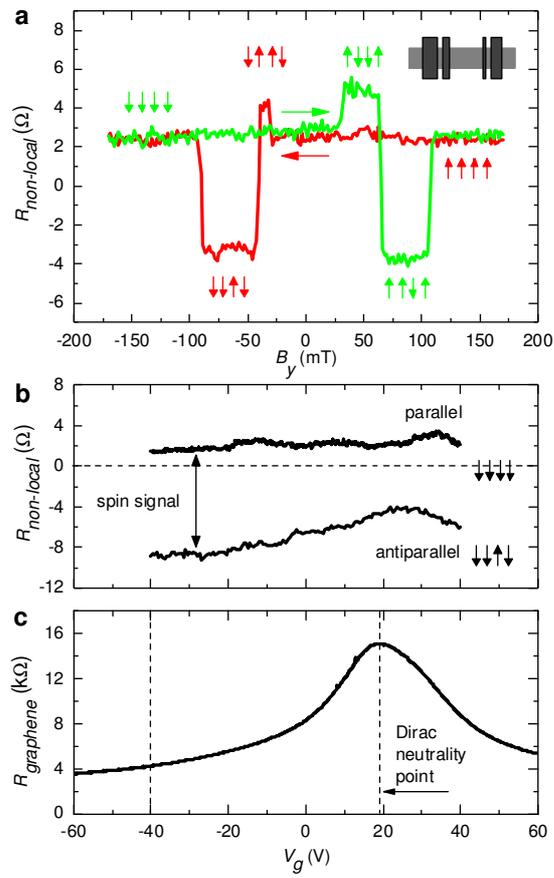

**Figure 4:**

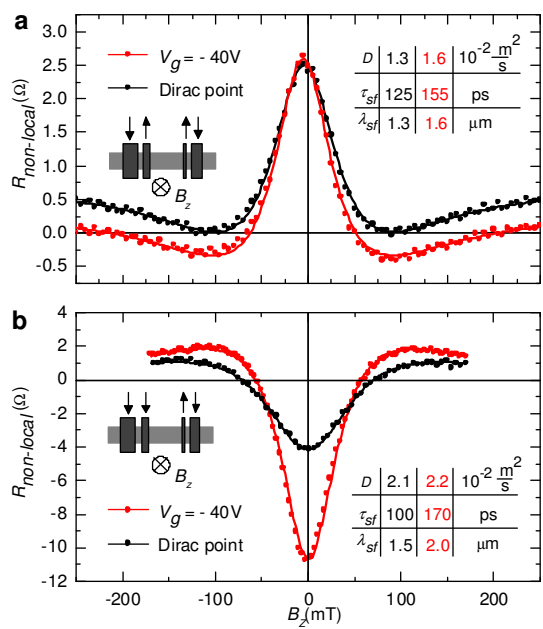

**Supplementary Methods**

**Measurements on devices with transparent low Ohmic contacts**

Here we give a brief overview of the results and conclusions on the devices with transparent and low Ohmic contacts. The preparation is similar to those described in the main text except for the Al evaporation and oxidation step. In two of these devices the ferromagnetic electrodes did not fully overlap the graphene flakes. As a result the graphene is exposed to stray magnetic fields in the range of several hundreds of mT at the ends of the strips. In four terminal measurements at 4.2 K we observed magnetoresistances of the order of 300 Ω on a background of 18 kΩ. These resistances coincided with the expected switching fields of the Co electrodes. Moreover the sign of the signals could change as function of the gate voltage but also as a function of the bias voltage. The sensitivity to magnetic field, charge density and energy strongly suggest that these are Hall type effects and we concluded that in order to avoid stray fields, the ferromagnets should completely overlap the graphene.

A third device had a geometry similar to device #1. Here we found (three terminal) contact resistances in the range < 50 Ω, therefore close to 0. This implies that there is a good contact between the ferromagnet and the graphene underneath, and the carriers are not able to pass underneath the contacts. This was confirmed by measurements of the non-local resistance yielding low values (<1 Ω). No spin valve signals could be detected. We also measured the two terminal local spin valve signal between contacts 2 and 3. On a background of typically 5 kΩ no signals could be observed, the noise level being about 5 Ω.

**Spin relaxation induced by the contacts**

We found the lowest contact resistances (devices #1,#2,#3,#4 and #5) to be around 5 kΩ, which corresponds to a resistance of about 10 kΩ per spin channel. This implies that a contact connects the spin-up and spin-down channels in the graphene by an effective resistance of 20 kΩ. Taking into account all 4 contacts, from this an effective relaxation time can be calculated due the spin relaxation via the contacts (see Zaffalon, M. & van Wees, B.J. Zero dimensional spin accumulation in a mesoscopic island. *Phys. Rev. Lett.* **91**, 186601 (2003)). An estimate shows that this time exceeds the observed spin relaxation time by at least a factor of 10. The contacts therefore contribute at most 10% to the observed spin relaxation.

**Experimental aspects**

i) The slight asymmetry in coercive fields of the ferromagnetic strips for positive and negative B sweeps arises from a "frozen-in" magnetization configuration in the wider parts of the ferromagnetic leads. Full symmetry can be restored by continuing the B sweeps to higher magnetic fields (~1T). In most cases this was not done for practical reasons.

ii) Non idealities in the non-local resistance measurements (in particular related to non-homogeneous contacts) can generate a spin independent background. This is estimated to be smaller than 1 Ω in typical measurements.